\documentclass[12pt]{iopart}   
\usepackage{graphicx}
\usepackage{epsfig}
\usepackage{amssymb}
\begin{document}

\newcommand{\be}{\begin{equation}}
\newcommand{\ee}{\end{equation}}
\newcommand{\beqn}{\begin{eqnarray}} 
\newcommand{\eeqn}{\end{eqnarray}}

\title{Infinite-disorder critical points of models with stretched exponential interactions}

\author{R\'obert Juh\'asz}
\address{Institute for Solid
State Physics and Optics, Wigner Research Centre for Physics, H-1525 Budapest,
P.O. Box 49, Hungary}
\ead{juhasz.robert@wigner.mta.hu}

\begin{abstract}
We show that an interaction decaying as a stretched exponential function of the distance, $J(l)\sim e^{-cl^a}$, is able to alter the universality class of short-range systems having an infinite-disorder critical point. 
To do so, we study the low-energy properties of the random transverse-field Ising chain with the above form of interaction 
by a strong-disorder renormalization group (SDRG) approach. 
We obtain that the critical behavior of the model is controlled by infinite-disorder fixed points different from that of the short-range one if $0<a<1/2$.
In this range, the critical exponents calculated analytically by a simplified SDRG scheme are found to vary with $a$, while, for $a>1/2$, the model belongs to the same universality class as its short-range variant. The entanglement entropy of a block of size $L$ increases logarithmically with $L$ in the critical point but, as opposed to the short-range model, the prefactor is disorder-dependent in the range $0<a<1/2$.  
Numerical results obtained by an improved SDRG scheme are found to be in agreement with the analytical predictions. 
The same fixed points are expected to describe the critical behavior of, among others, the random contact process with stretched exponentially decaying activation rates.  
\end{abstract}

\maketitle

\section{Introduction}

Long-range interactions are known to affect the cooperative behavior of many-particle systems in the vicinity of critical points. In the case that the strength of the interaction decays as a power $d+\sigma$ of the distance in $d$ dimensions, a frequently observed scenario is that, for a slow enough decay, i.e. for $\sigma<\sigma_{\rm MF}$, the critical behavior is of mean-field-like; for a sufficiently rapid decay, i.e. $\sigma>\sigma_{\rm SR}$, the phase transition falls into the universality class of the same model with short-range, i.e. finite-range interactions, while, in the intermediate regime $\sigma_{\rm MF}<\sigma<\sigma_{\rm SR}$, the critical exponents vary with $\sigma$.  
Another factor that can change the universality class of a phase transition is {\it quenched disorder}, which is inevitably present in many real systems \cite{bk,vojta_rev}.  
The description of the critical behavior of systems with long-range interactions in the presence of disorder is, in general, a hard problem, which is cumbersome to approach by analytical and even by numerical methods. 
Many works have been devoted to classical systems with the above two ingredients such as random-field ferromagnetic models \cite{rfim} or spin glasses \cite{sg} but, to quantum systems with long-range interactions with or even without disorder \cite{purelr}, less attention have been paid. 
Recently, the critical behavior of the zero-temperature random transverse-field Ising chain with algebraically decaying, long-range couplings has been studied \cite{jki}. 
The quantum critical behavior of the short-range variant of this model is known to be controlled by an {\it infinite-disorder fixed point} (IDFP) of a sequential, real-space renormalization procedure termed as strong-disorder renormalization group (SDRG) \cite{mdh,fisher,im}. 
The hallmark of an IDFP is an anisotropic relationship $\ln\tau\sim\xi^{\psi_{\rm SR}}$ between the time scale $\tau$ and length scale $\xi$, meaning that the dynamical exponent is formally infinite here.
The conclusion of the SDRG study of the long-range variant of the above model was that 
the critical behavior is controlled by a strong-disorder fixed point with a {\it finite} dynamical exponent $z=1+\sigma$ for any $\sigma>0$ \cite{jki}. 
Thus the scenario in (pure) long-range systems depicted above fails here, 
since the IDFP of the short-range model cannot be recovered no matter how large $\sigma$ is. 
In the above model, the form of the dynamical relationship $\tau\sim\xi^{d+\sigma}$ in the critical point seems to be dictated by the form of the distance-dependence of the interaction strength, $J(l)\sim l^{-(d+\sigma)}$, the inverse coupling corresponding to the time scale.  
This suggests that, even a more rapidly decreasing interaction, which 
decays as a stretched exponential, $J(l)\sim e^{-(l/l_0)^a}$, 
is able to change the dynamical relationship to $\ln\tau\sim\xi^a$
 if $a<\psi_{\rm SR}$, while it is expected to be irrelevant if $a>\psi_{\rm SR}$.  

The aim of this work is to study how the critical behavior of models having an IDFP is affected by the presence of a stretched exponentially decaying interaction. For concreteness, we will investigate the low-energy properties of the random transverse-field Ising chain by applying an SDRG approach, which, in its simplified form, provides analytical predictions on the large-scale behavior of the model. These will be shown to be in agreement with results of a more complete numerical SDRG scheme.  
According to the results, the critical behavior is controlled by an IDFP with 
$\psi=a$  for $0<a<\psi_{\rm SR}=1/2$. The scaling dimension of the average magnetization calculated analytically is found to vary with $a$, as well, and the entanglement entropy of a block of spins increases logarithmically with the size of the block with a disorder-dependent prefactor. 
The conclusions drawn here are expected to be generally valid for other models having an IDFP in their short-range form. 
In particular, we highlight a stochastic, nonequilibrium model, the {\it random contact process} \cite{cp,liggett,hiv}, the absorbing phase transition of which is expected to fall into the same universality class as the critical point of the random transverse-field Ising model.  

The structure of the paper is the following. In section \ref{sdrg}, an analytically tractable SDRG scheme of the model is presented. The properties of the fixed-point solution of the SDRG evolution equations are discussed in section \ref{fp}, and various physical quantities are calculated in section \ref{physical}. 
In section \ref{numerical}, the analytical predictions are compared against numerical results obtained by an improved SDRG scheme.  
Finally, the results are discussed in section \ref{discussion}.

\section{The SDRG method}
\label{sdrg}

Let us consider the random transverse-field Ising model (RTIM) defined by the Hamiltonian
\be
\mathcal{H}=-\sum_{i\neq j}J(l_{ij})\sigma^x_i\sigma^x_j - \sum_ih_i\sigma^z_i,
\label{hamiltonian}
\ee
where $\sigma^x_i$ and $\sigma^z_i$ are Pauli operators on site $i$.
The couplings depend on the distance $l_{ij}$ as 
\be 
J(l_{ij})=\omega_{ij}e^{-(l_{ij}/l_0)^a}
\label{Jl}
\ee
where $l_0$ and $a$ are positive constants, while the prefactors $\{\omega_{ij}\}$  and the transverse fields $\{h_i\}$ are i.i.d. positive, quenched random variables. 
For the sake of simplicity, the prefactors will be chosen to be non-random, 
$\omega_{ij}=J_0$, in the subsequent calculations; nevertheless, this variant of the model is expected to be in the same universality class as that with random prefactors. 
For a fixed $l_0$, $a$ and distribution $\rho(h)$ of the transverse fields, 
the prefactor $J_0$ can be used as a control parameter of the quantum phase transition of the model at zero temperature. 
For large enough $J_0$, the model is ferromagnetic with a positive average spontaneous magnetization $m_0=\lim_{H\to 0}\overline{\langle\sigma^x_0\rangle}(H)$, where $H$ is the magnitude of a magnetic field applied only on spin $0$ in the $x$ direction and the overbar denotes an average over disorder. 
At some critical value $J_0^c$, the model undergoes a quantum phase transition,
 and, for $J_0<J_0^c$, it will be paramagnetic with a vanishing spontaneous magnetization. 

In the calculations, we will restrict ourselves to the model in Eq. (\ref{hamiltonian}) in one dimension and assume that $0<a<\psi_{\rm SR}=1/2$. 
In the SDRG approach of the RTIM, the energy scale set by the largest coupling or transverse field $\Omega=\max\{J_{ij},h_i\}$ is gradually reduced by eliminating terms in the Hamiltonian with the largest parameter and calculating the remaining effective parameters perturbatively \cite{fisher,im}. 
Two kinds of reduction steps are applied iteratively. 
If the largest parameter is a coupling, $\Omega=J_{ij}$, 
the spins $i$ and $j$ form a spin cluster, which is subjected to an effective transverse field $\tilde h=h_ih_j/J_{ij}$ and has a magnetic moment 
$\tilde\mu=\mu_i+\mu_j$. 
If spin $i$ and $j$ had been coupled to another spin (cluster) $k$ before the decimation then the effective coupling of the cluster $ij$ to $k$ will be either
$\tilde J_{ij,k}=J_{ik}+J_{jk}$ if the so called ``sum rule'' is applied or 
 $\tilde J_{ij,k}=\max\{J_{ik},J_{jk}\}$ if the ``maximum rule'' is followed. 
If the largest parameter is a transverse field, $\Omega=h_i$, the spin cluster $i$ is decimated and new couplings between all pairs $(j,k)$ of spin clusters that were coupled to $i$ before the decimation are generated with the effective strength $\tilde J^0_{jk}=J_{ji}J_{ik}/h_i$. 
If a coupling $J_{jk}$ between cluster $j$ and $k$ existed before the decimation then, according to the sum rule, $\tilde J_{jk}=\tilde J^0_{jk}+J_{jk}$ whereas, according to the maximum rule, $\tilde J_{jk}=\max\{\tilde J^0_{jk},J_{jk}\}$. 
In an IDFP, where the distributions of logarithmic couplings and fields are broadening without limits, the SDRG approach both with the sum rule and the maximum rule becomes asymptotically exact and is conjectured to provide the correct critical exponents \cite{fisher,im}.  

First, we shall consider the SDRG scheme with the maximum rule, which, by further simplifying assumptions valid close to the fixed point in the spirit of Ref. \cite{jki}, reduces to an analytically tractable scheme.
Let us investigate what consequences the maximum rule has in the model under study. When a bond $J_{i,i+1}$ connecting the adjacent clusters $i$ and $i+1$ is decimated, the effective field of the new cluster formed from them will be  
$\tilde h=h_ih_{i+1}/J_{i,i+1}$ and the effective coupling of this cluster to  
clusters $k>i+1$ ($k<i$) will be given by the couplings $J_{i+1,k}$ ($J_{k,i}$) 
of its constituent $i+1$ ($i$) before the decimation.
If a transverse field $h_i$ is decimated, the indirect coupling 
$\tilde J^0_{jk}=J_{ji}J_{ik}/h_i$ through the decimated cluster $i$ between $j$ and $k$ has to be compared to the existing long-range coupling $J(l_{jk})$.
According to numerical SDRG investigations, the long-range couplings almost always exceed the indirect ones as the critical fixed point is approached. 
This can be understood intuitively, since, in the critical short-range model, the effective couplings between adjacent clusters decrease with the spacing $l$ between them as $J(l)\sim\exp(-cl^{\psi_{\rm SR}})$, i.e. more rapidly than long-range couplings in Eq. (\ref{Jl}) for $a<\psi_{\rm SR}=1/2$.  
On the basis of this observation, we postulate that the renormalized coupling between clusters $i-1$ and $i+1$ is 
\be 
\tilde J_{i-1,i+1}=J(l_{i-1,i+1}),
\ee
where the distance $l_{ij}$ between clusters is defined as the distance between their closest constituent spins, while other couplings remain unchanged when $h_i$ is decimated. 
Applying these simplified decimation rules, one can see that, at any stadium of the SDRG procedure, the coupling between any pairs of clusters will be given by the long-range coupling between their closest spins, and exclusively couplings between neighboring clusters are chosen for decimation. 
If a cluster having a transverse field $\Omega=h_i$ is eliminated, it is expedient to formulate the decimation rule in terms of length variables 
\be
l_{ij}\equiv l_0\left[\ln\left(\frac{J_0}{J_{ij}}\right)\right]^{1/a}
\ee
rather than in terms of couplings as 
\be 
\tilde l_{i-1,i+1}=l_{i-1,i}+l_{i,i+1}+w_i,
\ee
where $w_i$ denotes the distance between the end spins of cluster $i$. 
We can see that, within the simplified scheme, besides the transverse field $h_i$ and length $w_i$ of clusters, it is sufficient to the keep track of the 
couplings or, equivalently, the distances $l_{i,i+1}$ between {\it neighboring} clusters only. In this respect, the SDRG scheme is similar to that of the one-dimensional short-range model, although the decimation rules are different. 

Before turning to the analysis of the above scheme, a caveat is in order concerning its validity. 
The application of the maximum rule, which leads to that only the interactions between closest spins of clusters are kept, is justified by that 
the interaction strength decreases rapidly with the distance. 
In the paramagnetic phase and in the critical point, where the spin clusters produced by the method are sparse, this approximation is expected to 
be reasonable. In the ferromagnetic phase, however, where the SDRG method produces large, compact clusters, its reliability is questionable, therefore we will not analyze it in that phase. 

To write the decimation rules in an additive form, let us introduce the logarithmic energy scale 
\be
\Gamma=\ln(J_0/\Omega),
\ee 
and the reduced variables 
\be 
\beta_i=\ln(\Omega/h_i), \qquad 
\chi_i=\frac{w_i}{l_0\Gamma^{1/a}}, \qquad 
\zeta_i=\left[\frac{\ln(J_0/J_{i,i+1})}{\Gamma}\right]^{1/a}-1,
\ee
which lie in the range $[0,\infty)$ as soon as $\Omega$ is reduced below $J_0$ \footnote{Note that, as $l_{i,i+1}$ and $w_i$ take on integer values, the variables $\zeta$ and $\chi$ are discrete for any finite $\Gamma$ but become quasicontinuous in the limit $\Gamma\to\infty$. For the sake of simplicity, we will treat them as continuous variables in the followings.}. 
In terms of these variables, the transformation rule for the decimation of a coupling $\Omega=J_{i,i+1}$ takes the simple form
\beqn 
\tilde\beta=\beta_i+\beta_{i+1}, 
\label{b_rule}
\\
\tilde\chi=\chi_i+\chi_{i+1}+\zeta_i+1,
\eeqn
while, if a transverse field $\Omega=h_i$ is decimated, we have 
\be 
\tilde\zeta=\zeta_{i-1}+\zeta_i+\chi_i+1.
\label{z_rule}
\ee
We can see that the variables $\beta_i$ and $\zeta_j$ on different places remain independent during the SDRG procedure, while $\beta_i$ and $\chi_i$ become correlated. 
Thus, for a complete characterization of the model, one should follow up the evolution of the probability density $f_{\Gamma}(\zeta)$ and that of the joint probability density $p_{\Gamma}(\beta,\chi)$ as the parameter $\Gamma$ increases during the SDRG procedure. 
Instead of this full problem, which is difficult to treat, we concentrate on the evolution of 
$f_{\Gamma}(\zeta)$ and that of the marginal probability density 
$g_{\Gamma}(\beta)$, and will see that many asymptotic properties of the model can be extracted from these functions.
 The evolution equations of these distributions under the progression of the SDRG procedure can be derived in a standard way and read as 
\beqn
\frac{\partial g_{\Gamma}(\beta)}{\partial\Gamma}=
\frac{\partial g_{\Gamma}(\beta)}{\partial\beta} + \nonumber \\
+\frac{f_0}{a\Gamma}\int d\beta_1\int d\beta_2g_{\Gamma}(\beta_1)g_{\Gamma}(\beta_2)\delta(\beta-\beta_1-\beta_2) +g_{\Gamma}(\beta)\left(g_0-\frac{f_0}{a\Gamma}\right) 
\label{g_ev}
\\
\frac{\partial f_{\Gamma}(\zeta)}{\partial\Gamma}=
\frac{\zeta+1}{a\Gamma}\frac{\partial f_{\Gamma}(\zeta)}{\partial\zeta}+ \nonumber \\ 
+g_0\int d\zeta_1\int d\zeta_2\int d\chi f_{\Gamma}(\zeta_1)f_{\Gamma}(\zeta_2)q_{\Gamma}(\chi)\delta(\zeta-\zeta_1-\zeta_2-\chi-1)+
f_{\Gamma}(\zeta)\left(\frac{f_0+1}{a\Gamma}-g_0\right), \nonumber \\
\label{f_ev}
\eeqn
where $g_0(\Gamma)\equiv g_{\Gamma}(0)$, $f_0(\Gamma)\equiv f_{\Gamma}(0)$,  
$q_{\Gamma}(\chi)\equiv p_{\Gamma}(0,\chi)/g_{\Gamma}(0)$ is the conditional probability density of $\chi$ given the occurrence of $\beta=0$,  
and $\delta$ denotes the Dirac delta function. 
The first terms on the r.h.s. of the equations appear owing to that $\beta$ and $\zeta$ depend on $\Gamma$; the integrals are related to the generation of effective parameters according to the rules in Eqs. (\ref{b_rule}) and (\ref{z_rule}), while the last terms ensure the normalization of the distributions.  

\section{Properties of the fixed-point solution}
\label{fp}

\subsection{Paramagnetic phase}

The fixed-point solution of Eq. (\ref{g_ev}) is of the form 
\be
g_{\Gamma}(\beta)=g_0(\Gamma)\exp[-g_0(\Gamma)\beta],
\label{gfp}
\ee
which, by substituting into Eq. (\ref{g_ev}),  
leads to the differential equation 
\be 
\frac{dg_0(\Gamma)}{d\Gamma}=-\frac{f_0(\Gamma)g_0(\Gamma)}{a\Gamma}.
\label{g_diff}
\ee
for the functions $g_0(\Gamma)$ and $f_0(\Gamma)$.
Let us first consider the paramagnetic phase of the model. We will see {\it a posteriori} that $f_0(\Gamma)$ tends to zero here in the limit $\Gamma\to\infty$ and, accordingly, the typical $\zeta$ will increase without limits.  
Moreover, since the ratio $r(\Gamma)$ of the frequency of bond and field decimations tends to zero, the typical distance between adjacent clusters will increase much faster than their typical extension. Therefore, typically $\zeta\gg \chi$, and the delta function in Eq. (\ref{f_ev}) can be replaced by $\delta(\zeta-\zeta_1-\zeta_2)$ close to the fixed point.  
The fixed-point solution of this simplified equation is 
$f_{\Gamma}(\zeta)=f_0(\Gamma)\exp[-f_0(\Gamma)\zeta]$, provided that
$f_0(\Gamma)$ satisfies the differential equation 
\be 
\frac{df_0(\Gamma)}{d\Gamma}=f_0(\Gamma)\left[\frac{1}{a\Gamma}-g_0(\Gamma)\right]. 
\label{f_diff}
\ee
The functional form $f_0(g_0)$ of the renormalization group trajectories in the $f_0-g_0$ plane can be derived as follows. Using the function $G(\Gamma)\equiv \Gamma g_0(\Gamma)$, instead of $g_0(\Gamma)$, the above differential equations can be reformulated as  
\beqn
\frac{dG(\Gamma)}{d\Gamma}=\frac{G(\Gamma)}{\Gamma}\left(1-\frac{f_0(\Gamma)}{a}\right), \nonumber \\
\frac{df_0(\Gamma)}{d\Gamma}=\frac{f_0(\Gamma)}{\Gamma}\left(\frac{1}{a}-G(\Gamma)\right).
\eeqn
Eliminating $\Gamma$ from these equations, we obtain 
\be 
\frac{df_0}{dG}=\frac{f_0(1/a-G)}{G(1-f_0/a)}, 
\ee
which, after integration, results in 
\be 
f_0e^{-f_0/a}=G^{1/a}e^{-G}C
\label{integral} 
\ee
with a constant $C$. This relation can be recast in an explicit form with respect to $f_0$ by the help of the Lambert $W$ function
\be 
f_0=-aW(-a^{-1}G^{1/a}e^{-G}C).
\label{lambert}
\ee
The trajectories in the paramagnetic phase have the limits $g_0(\Gamma)\to {\rm const}=\delta>0$ and $f_0(\Gamma)\to 0$ for $\Gamma\to\infty$. Then, according to Eq. (\ref{lambert}),  $f_0(\Gamma)$ tends to zero asymptotically as 
\be 
f_0(\Gamma)=-aW(-a^{-1}(\Gamma g_0)^{1/a}e^{-\Gamma g_0}C)\simeq C(\Gamma\delta)^{1/a}e^{-\Gamma\delta}.  
\label{f_gamma}
\ee
The ratio of the frequency of bond and field decimations thus vanishes as 
\be 
r(\Gamma)=\frac{f_0(\Gamma)}{a\Gamma g_0(\Gamma)}
\sim (\Gamma\delta)^{1/a-1}e^{-\Gamma\delta} 
\ee
for large $\Gamma$. 
To the constant $\delta$, a physical meaning can be assigned by deriving
a relationship between the energy scale $\Omega$ and the length scale $\ell=1/n$, where $n$ is the mean number of active (non-decimated) spin clusters per unit length of the chain. 
The function $n(\Gamma)$ obeys the differential equation 
\be 
\frac{dn(\Gamma)}{d\Gamma}=-n(\Gamma)\left[g_0(\Gamma)+\frac{f_0(\Gamma)}{a\Gamma}\right],
\label{n_gamma}
\ee
and, using that $g_0(\Gamma)+f_0(\Gamma)/a\Gamma\simeq\delta$ for large $\Gamma$, we obtain 
\be
\ell(\Gamma)\simeq Ce^{\delta\Gamma}=C\left(\frac{J_0}{\Omega}\right)^{\delta}
\ee
with a constant of integration $C$. 
Thus, the constant $\delta$ can be interpreted as the inverse of the dynamical exponent: 
\be 
\delta=1/z.
\ee
The lowest energy gap $\epsilon_L$ of a large but finite system of size $L$ is given in the paramagnetic phase by the twice of the effective transverse field of the last decimated cluster. 
From Eq. (\ref{gfp}), we obtain that the distribution of transverse fields is a power law,
$p(h)\simeq\frac{1}{z\Omega}(h/\Omega)^{-1+1/z}$, close to the fixed point.  
Therefore the energy gap, being the twice of the smallest one among $O(L)$ transverse fields, will follow a Fr\'echet distribution and scale as $\epsilon_L\sim L^{-z}$ for large $L$, according to extreme value statistics \cite{galambos,jli}. 
Thus, the system is gapless in this phase due to the occurrence of ferromagnetic clusters of unbounded size and expected to show Griffiths-McCoy singularities analogous to the short-range model \cite{griffiths,vojta_rev}.

\subsection{Critical point}

In the limit $\delta\to 0$, the critical point of the model is approached, and, in that point, the form of the functions $g_0(\Gamma)$ and $f_0(\Gamma)$ will be different from those in the paramagnetic phase. 
Considering still the simplified equation, the limiting values will be 
$g_0(\Gamma)\to 0$ and $f_0(\Gamma)\to a$. Thus, $f_{\infty}\equiv\lim_{\Gamma\to\infty}f_0(\Gamma)$ being non-zero, the neglections in the argument of the delta function are not justified in the critical point and, consequently, the fixed-point distribution $f_{\Gamma}(\zeta)$ is not a pure exponential here. 
Nevertheless, according to numerical investigations of the SDRG procedure, 
the simplified equation predicts qualitatively correctly that the limiting value $f_{\infty}$ is finite and positive.  
With this assumption, Eq. (\ref{g_diff}) then gives the leading order $\Gamma$-dependence of $g_0(\Gamma)$ in the form $g_0(\Gamma)\simeq b/\Gamma^{\alpha}$ 
with $\alpha=f_{\infty}/a$ and an unknown positive constant $b$. 
Let us first consider the possibility $\alpha>1$. Then the terms proportional to $g_0(\Gamma)$ in Eq. (\ref{f_ev}) can be neglected and the resulting equation would have the fixed-point solution  $f_{\Gamma}(\zeta)=f_0(\Gamma)\exp[-f_0(\Gamma)\zeta]$ with $f_0(\Gamma)$ obeying 
$\frac{df_0(\Gamma)}{d\Gamma}=\frac{f_0(\Gamma)}{a\Gamma}$. But this gives $f_0(\Gamma)\sim \Gamma^{1/a}$, which is in contradiction with the assumption that  
$f_{\infty}$ is finite. 
Now, let us consider the possibility $\alpha<1$ and restrict ourselves to the range $0\le\zeta\le 1$, where the integral on the r.h.s. of Eq. (\ref{f_ev}) is identically zero. Then, keeping the leading order terms in $1/\Gamma$,  Eq. (\ref{f_ev}) reduces in this range to 
$\frac{\partial f_{\Gamma}(\zeta)}{\partial\Gamma}\simeq -g_0(\Gamma)f_{\Gamma}(\zeta)$. This leads to $f_{\infty}=0$, again in contradiction with the assumption $f_{\infty}>0$. 
We thus conclude that $\alpha=1$ or, equivalently, $f_{\infty}=a$. 
In order to determine the unknown constant $b$, one should be able to treat the full problem of the evolution of the functions $p_{\Gamma}(\beta,\chi)$ and 
$f_{\Gamma}(\zeta)$. 
But fortunately, there is a simple, alternative way of finding $b$. 
Substituting the expressions $g_{0}(\Gamma)\simeq b/\Gamma$ and  $f_{0}(\Gamma)\simeq a$ into Eq.(\ref{n_gamma}), one obtains the relationship 
\be
\ell(\Gamma)\simeq (\Gamma/\Gamma_0)^{1+b},
\label{dyn_crit}
\ee 
where the constant $\Gamma_0$ depends on the initial parameters of the model. 
On the other hand, the length scale $\ell(\Gamma)$ is asymptotically proportional to the average distance $\overline{l}(\Gamma)$ between adjacent clusters. Using that the distribution of the scaling variable $\zeta=l/l_0\Gamma^{1/a}-1$ converges to a $\Gamma$-independent limit distribution, we conclude that $\ell(\Gamma)\sim\overline{l}(\Gamma)\sim\Gamma^{1/a}$. 
Comparing this relation to Eq. (\ref{dyn_crit}), we obtain 
\be
b=(1-a)/a.
\ee 
The ratio of the frequency of bond and field decimations will thus tend to a finite limit 
\be 
r(\Gamma)=\frac{f_0(\Gamma)}{a\Gamma g_0(\Gamma)}\to \frac{a}{1-a},
\ee
which is different from one, indicating that the duality between the couplings and transverse fields characteristic of the critical fixed point of the short-range model \cite{fisher} is broken here. 
We will see in the subsequent sections that, knowing the leading terms of the functions $g_{0}(\Gamma)$ and $f_{0}(\Gamma)$ for large $\Gamma$, several asymptotic properties of the model can be calculated. 

Let us now turn to the question of how the correlation length $\xi$ of the average spatial correlations of the operator $\sigma_i^x$ diverges as the critical point is approached in the paramagnetic phase, and use the inverse dynamical exponent $\delta$ as a quantum control parameter. In the paramagnetic phase, at some $\delta$, the average length of spin clusters, which is of the same order of magnitude as the correlation length $\xi_{\delta}$, is finite, and, in the SDRG procedure, essentially no coupling decimations occur beyond this scale, $\ell\gg \xi_{\delta}$. 
Close to the critical point, $\delta\ll 1$, the logarithmic energy scale corresponding to $\xi_{\delta}$ is $\Gamma_{\delta}\sim \xi_{\delta}^a$. 
As can be observed e.g. in Eq. (\ref{f_gamma}), the solutions contain the scaling combination $\Gamma\delta$, suggesting the relation $\Gamma_{\delta}\sim \delta^{-1}$ close to the critical point. This yields for the divergence of the correlation length 
\be 
\xi_{\delta}\sim \delta^{-\nu}
\ee
with the correlation-length exponent 
\be
\nu=1/a.
\ee

\section{Scaling of physical quantities in the critical point}
\label{physical}

\subsection{Magnetization}

Within the SDRG approach, the scaling of the average spontaneous magnetization $m_0$ can be determined from that of the probability $\mathcal{S}(\Gamma)$ that a given spin is part of an active cluster at the logarithmic energy scale $\Gamma$. 
In order to obtain $\mathcal{S}(\Gamma)$, let us consider the probability $s_{\Gamma}(\beta)d\beta$ that, at the scale $\Gamma$, a given spin is part of an active cluster with a logarithmic transverse field $\beta$. 
Similarly to Eqs. (\ref{g_ev}-\ref{f_ev}), we can formulate an evolution equation for $s_{\Gamma}(\beta)$ under the change of $\Gamma$, which reads as 
\be 
\frac{\partial s_{\Gamma}(\beta)}{\partial\Gamma}=
\frac{\partial s_{\Gamma}(\beta)}{\partial\beta} - \frac{2f_0(\Gamma)}{a\Gamma}
\left[s_{\Gamma}(\beta) - \int_0^{\beta}d\beta's_{\Gamma}(\beta')g_{\Gamma}(\beta-\beta')\right].
\label{s_ev}
\ee
The solution of such an equation can be found by the ansatz \cite{rm}
\be 
s_{\Gamma}(\beta)=[u(\Gamma)+v(\Gamma)g_0(\Gamma)\beta]
g_0(\Gamma)e^{-g_0(\Gamma)\beta}, 
\label{ansatz}
\ee
where $u(\Gamma)$ and $v(\Gamma)$ are unknown functions of $\Gamma$. 
Substituting this and the fixed-point solution $g_{\Gamma}(\beta)$ in Eq. (\ref{gfp}) into Eq. (\ref{s_ev}), it turns out to be a solution, indeed, provided that $u(\Gamma)$ and $v(\Gamma)$ satisfy the differential equations
\beqn 
\frac{du(\Gamma)}{d\Gamma}=-\left[g_0(\Gamma)+\frac{f_0(\Gamma)}{a\Gamma}\right]u(\Gamma)+g_0(\Gamma)v(\Gamma)  \nonumber \\
\frac{dv(\Gamma)}{d\Gamma}=\frac{f_0(\Gamma)}{a\Gamma}u(\Gamma)-g_0(\Gamma)v(\Gamma). 
\eeqn
Using now the asymptotic forms $g_0(\Gamma)\simeq \frac{1-a}{a\Gamma}$ and $f_0(\Gamma)\simeq a$ valid in the critical point for large $\Gamma$, these can be rewritten as linear differential equations with constant coefficients
\beqn 
\frac{du}{d\gamma}\simeq -\left(\frac{1-a}{a}+1\right)u+\frac{1-a}{a}v  \nonumber \\
\frac{dv}{d\gamma}\simeq u -\frac{1-a}{a}v,  
\eeqn 
in terms of the new independent variable $\gamma\equiv\ln\Gamma$. 
The solutions of these equations decay with $\Gamma$ asymptotically as 
$u(\Gamma)\sim v(\Gamma)\sim e^{\gamma\epsilon_+}\sim \Gamma^{\epsilon_+}$, where 
$\epsilon_+=-\frac{1-a}{a}-\frac{1}{2}+\sqrt{\frac{1-a}{a}+\frac{1}{4}}$ is the larger eigenvalue of the coefficient matrix. 
The probability $\mathcal{S}(\Gamma)$ we are looking for can be expressed by these functions as
\be 
\mathcal{S}(\Gamma)=\int_0^{\infty}s_{\Gamma}(\beta)d\beta=u(\Gamma)+v(\Gamma),  
\ee
and thus scales as $\mathcal{S}(\Gamma)\sim\Gamma^{\epsilon_+}$. 
This yields for the scaling of the average spontaneous magnetization with the size $L$ of finite systems, using $\Gamma\sim L^a$, 
\be
m_0(L)\sim L^{-x} 
\label{mL}
\ee
with the scaling dimension 
\be 
x=a|\epsilon_+|=a\left(\frac{1-a}{a}+\frac{1}{2}-\sqrt{\frac{1-a}{a}+\frac{1}{4}}\right).
\label{x}
\ee

\subsection{Surface magnetization}

Next, let us consider the average spontaneous magnetization $m_0^s$ at the end spin of semi-infinite chains. 
Concerning the end spin, the evolution of the corresponding function $s_{\Gamma}(\beta)$ is governed by the equation
\be 
\frac{\partial s_{\Gamma}(\beta)}{\partial\Gamma}=
\frac{\partial s_{\Gamma}(\beta)}{\partial\beta} - \frac{f_0(\Gamma)}{a\Gamma}
\left[s_{\Gamma}(\beta) - \int_0^{\beta}d\beta's_{\Gamma}(\beta')g_{\Gamma}(\beta-\beta')\right],
\label{s_ev_surf}
\ee
which differs from Eq. (\ref{s_ev}) in that the factor $2$ is absent in front of the 2nd term on the r.h.s. since the end spin has only one neighbor. 
The fixed-point solution of this equation has the form 
\be 
s_{\Gamma}(\beta)=\mathcal{S}(\Gamma)g_0(\Gamma)e^{-g_0(\Gamma)\beta}. 
\ee
Substituting this and the fixed-point distribution $g_{\Gamma}(\beta)$ in Eq. (\ref{gfp}) into Eq. (\ref{s_ev_surf}) results in
\be 
\frac{d\mathcal{S}(\Gamma)}{d\Gamma}=-g_0(\Gamma)\mathcal{S}(\Gamma).
\ee
Using the asymptotic form $g_0(\Gamma)\simeq \frac{1-a}{a\Gamma}$ valid in the critical point, we obtain ultimately
\be 
\mathcal{S}(\Gamma)\simeq \left(\Gamma/\Gamma_0\right)^{-(1-a)/a}.
\ee
This yields for the finite-size scaling of the surface magnetization 
\be 
m_0^s(L)\sim L^{-x_s}
\ee
with the surface scaling dimension 
\be 
x_s=1-a.
\ee

\subsection{Entanglement entropy}

The entanglement entropy of critical quantum systems has attracted much interest recently \cite{afov}. 
Let us assume that the model under study is in its ground-state $|0\rangle$ and consider a block of $L$ contiguous spins in it; the rest of the (infinitely large) total system is referred to as the environment.   
The entanglement entropy $S_L$ is the von Neumann entropy of the reduced density operator $\rho_L={\rm Tr}_{E}|0\rangle\langle 0|$ of the block, where ${\rm Tr}_E$ denotes a partial trace over the environment: 
\be 
S_L=-{\rm Tr}(\rho_L\ln\rho_L).
\ee 
Within the SDRG approach, the entanglement entropy of a block is given by the number of clusters containing spins both inside and outside the block, multiplied by $\ln 2$ \cite{rm}. 
The asymptotic dependence of the average entanglement entropy $\overline{S_L}$ on $L$ can be obtained in the following way.
Let us assume that the model is renormalized down to $\Gamma=0$, so that, ultimately, all spins of the original model are organized into clusters, which were decimated at some $\Gamma$. 
Consider then the length $l$ of spacings between neighboring spins belonging to the same cluster and let 
$\mathcal{N}(l)$ denote the mean number of the occurrence of the distance $l$ per unit length of the chain. 
Obviously, a spacing of length $l$ appears in a cluster whenever a bond of length $l$ is decimated during the SDRG procedure, and this event occurs precisely at the scale 
\be 
\Gamma=(l/l_0)^a
\label{Gl}
\ee 
since the strength and the length of a bond are connected according to Eq. (\ref{Jl}).
When the mean concentration of clusters $n$  is reduced to $n+dn$ during the procedure, the mean number of bond decimations per unit length of the chain will be
$|dn|$ multiplied by the relative frequency $\frac{f_0/a\Gamma}{g_0+f_0/a\Gamma}$ of bond decimations.
Approximating $l$ by a continuous variable, we can write 
\beqn 
\frac{f_0/a\Gamma}{g_0+f_0/a\Gamma}|dn|=\frac{f_0/a\Gamma}{g_0+f_0/a\Gamma}\left|\frac{dn}{d\Gamma}\right|\frac{d\Gamma}{dl}dl= \nonumber \\
=n[\Gamma(l)]\frac{f_0}{a\Gamma}a\Gamma l^{-1}dl=
n[\Gamma(l)]f_0[\Gamma(l)]l^{-1}dl,
\eeqn
where we have used Eq. (\ref{n_gamma}) and Eq. (\ref{Gl}).
For large $l$, we have thus $\mathcal{N}(l)\simeq n[\Gamma(l)]f_0[\Gamma(l)]l^{-1}$.
The fraction of the chain covered by bonds of length $l$ is given by 
$l\mathcal{N}(l)$ and has the asymptotic form for large $l$
\be 
l\mathcal{N}(l)\simeq n[\Gamma(l)]f_0[\Gamma(l)]\simeq a[\Gamma(l)/\Gamma_0]^{-1/a}=a\Gamma_0^{1/a}\frac{l_0}{l},
\ee
where we have used Eq. (\ref{dyn_crit}). 
The large-$L$ asymptotics of the average entanglement entropy can be obtained 
from $\mathcal{N}(l)$ as 
\be 
\overline{S_L}/\ln 2\simeq 2\int^Ll\mathcal{N}(l)dl + 
2L\int_L^{\infty}\mathcal{N}(l)dl\simeq 2a\Gamma_0^{1/a}l_0\ln L + {\rm const}.
\ee

\section{Numerical SDRG analysis}
\label{numerical}

We have performed a numerical investigation of the model by an improved SDRG scheme. This differs from the one treated analytically in that it takes into account the long-range interaction between all pairs of spins of adjacent clusters by using the sum rule. 
To be concrete, the effective interactions between neighboring clusters are described by a single coupling $J_{n,n+1}$ as before, but the decimation rules are modified as follows. 
If a coupling $J_{n,n+1}$ is decimated, the effective transverse field of the new cluster $\mathcal{C}_{\tilde n}=\mathcal{C}_n\cup\mathcal{C}_{n+1}$ is calculated as  $\tilde h=h_nh_{n+1}/J_{n,n+1}$ but, at the same time, the coupling of $\mathcal{C}_{\tilde n}$ to $\mathcal{C}_{n+2}$ will be modified to
\be 
J_{\tilde n,n+2}=J_{n+1,n+2}+\sum_{i\in \mathcal{C}_n, j\in \mathcal{C}_{n+2}}J(l_{ij}),
\ee
and, similarly, the coupling to $\mathcal{C}_{n-1}$ will be $J_{n-1,\tilde n}=J_{n-1,n}+\sum_{i\in \mathcal{C}_{n-1}, j\in \mathcal{C}_{n+1}}J(l_{ij})$.
If the transverse field $h_n$ is decimated, the effective coupling between 
cluster $\mathcal{C}_{n-1}$ and $\mathcal{C}_{n+1}$ will be 
\be
\tilde J_{n-1,n+1}=\frac{J_{n-1,n}J_{n,n+1}}{h_n} + 
\sum_{i\in \mathcal{C}_{n-1}, j\in \mathcal{C}_{n+1}}J(l_{ij}).
\ee

We have implemented the above decimation rules numerically, starting with a uniform distribution of the transverse fields in the range $[0,1]$, and fixing the parameters in Eq. (\ref{Jl}) to $a=1/3$ and $l_0^{-a}=2$, while used $J_0$ as a control parameter. After renormalizing rings of $L=2^4-2^{18}$ spins down to $2$ spin clusters, averages of different quantities over $10^5$ independent realizations of the disorder have been calculated.

\begin{figure}[h]
\includegraphics[width=0.7\linewidth]{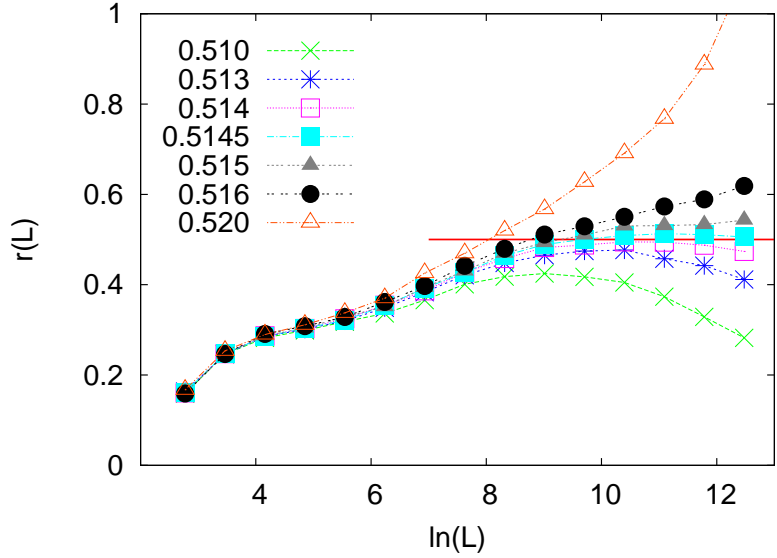}
\caption{
\label{rp2}
The ratio of the frequency of coupling and field decimations in the last decimation step plotted against $\ln L$ for different values of the control parameter $J_0$. 
The horizontal line indicates the limiting value $1/2$ obtained by the simplified SDRG scheme for $a=1/3$.  
}
\end{figure}

First, we considered the ratio $r(L)$ of the frequency of coupling and field decimations in the last decimation step as a function of $L$ for different $J_0$.
As can be seen in Fig. \ref{rp2}, for small (large) values of $J_0$, $r(L)$
decreases (increases) for large $L$, while, at a critical value $J_0^c=0.5145(5)$, it seems to tend to a constant $0.51(2)$, which is close to the limiting value $a/(1-a)=1/2$ predicted by the simplified SDRG scheme.  

\begin{figure}[h]
\includegraphics[width=0.7\linewidth]{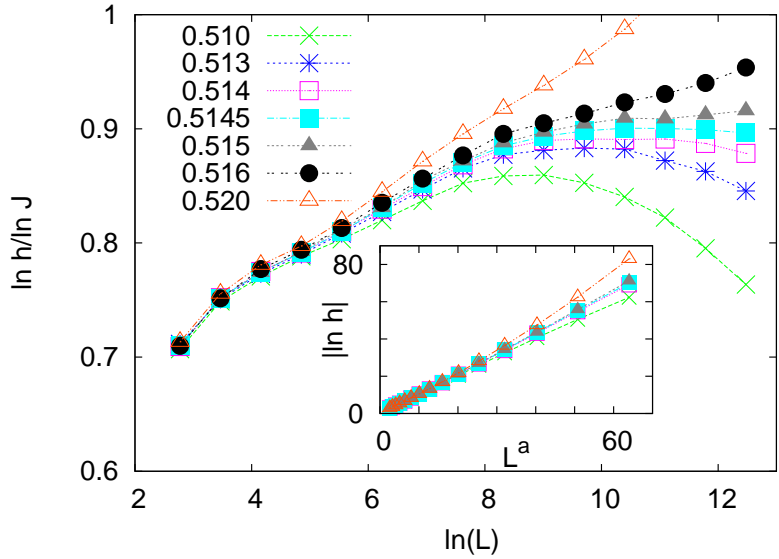}
\caption{
\label{gapp2}
The ratio $\overline{\ln h(L)}/\overline{\ln J(L)}$
of average logarithmic parameters of the renormalized system plotted against $\ln L$ for different values of $J_0$. 
The inset shows the average logarithmic transverse field $|\overline{\ln h(L)}|$ as a function of $L^a$. 
}
\end{figure}

We have also calculated the average of the logarithm of the effective transverse fields and couplings of the renormalized rings 
for different initial sizes $L$. 
According to the analytical results, the  averages of logarithmic parameters scale with $\Gamma$ asymptotically as 
$-\overline{\ln h}\sim -\overline{\ln J}\sim\Gamma$ in the critical point. 
The dynamical relation $\Gamma\sim L^a$ then yields
the finite-size scaling law
\be 
-\overline{\ln h(L)} \sim -\overline{\ln J(L)} \sim L^a.
\label{fss}
\ee
As can bee seen in Fig. \ref{gapp2}, the ratio $\overline{\ln h(L)}/\overline{\ln J(L)}$ tends to a constant in the estimated critical point, and the finite-size scaling behavior of $\overline{\ln h(L)}$ for large $L$ is compatible with the form given in Eq. (\ref{fss}). 

\begin{figure}[h]
\includegraphics[width=0.7\linewidth]{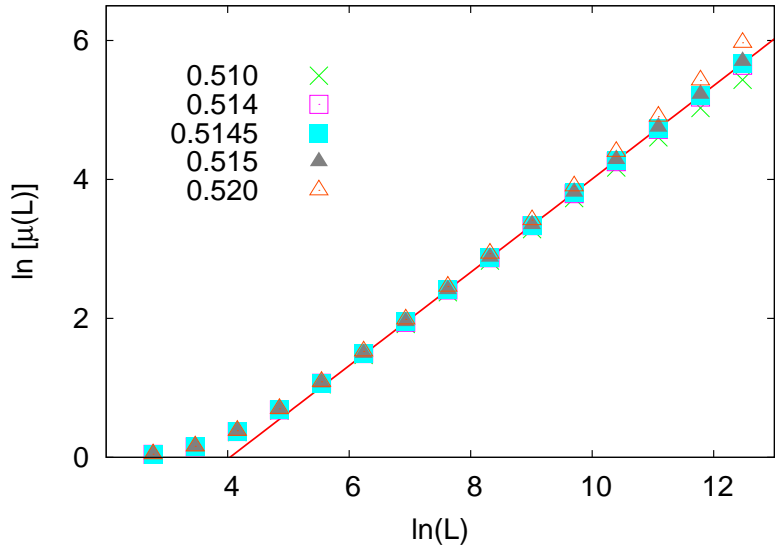}
\caption{
\label{mp2}
The logarithm of the average magnetic moment of clusters in the renormalized rings plotted against $\ln L$ for different values of $J_0$. The straight line is a linear fit to the data obtained in the estimated critical point and has the slope $d_f=0.67(1)$.
}
\end{figure}

Finally, the average magnetic moment $\overline{\mu}(L)$ of clusters in the renormalized rings has been calculated for different $L$, as well. 
According to the analytical results for the spontaneous magnetization in Eq. (\ref{mL}), the average magnetic moment in the critical point is expected to scale as 
\be 
\overline{\mu}(L)\sim m_0(L)L \sim L^{d_f},
\ee
where $d_f=1-x$ is the fractal dimension of clusters. 
Using the result in Eq. (\ref{x}), the fractal dimension for $a=1/3$ is $d_f=2/3$. 
The size-dependence of the average magnetic moment is shown in Fig. \ref{mp2} in a log-log plot. A linear fit to the data obtained in the critical point gives an estimate for the fractal dimension $d_f=0.67(1)$, which is compatible with the prediction provided by the simplified SDRG method.   

\section{Discussion}
\label{discussion}

We have studied in this work the low-energy properties of the random transverse-field Ising chain with long-range interactions that decay as a stretched exponential function the distance. 
Applying an SDRG procedure, analytical results have been obtained in the frame of a simplified scheme based on the maximum rule, and these were found to be compatible with numerical results obtained by an improved scheme that partially works with the sum rule. 
From a technical point of view, the main difference between the SDRG scheme of the long-range model and that of the short-range one is that there is one less ``free'' parameter here, as the strength $J$ and length $l$ of bonds, which are not perfectly correlated in the short-range model, are connected here through the relation $J=J_0e^{-(l/l_0)^a}$.
According to the results, the critical behavior is controlled by an IDFP with a generalized dynamical exponent $\psi=a$, and the other critical exponents $x$ and $\nu$ are found to vary with $a$ in the range $0<a<1/2$, as well.
Taking the limit $a\to \psi_{\rm SR}=1/2$, the above critical exponents agree with those of the short-range random transverse-field Ising chain, and, above the threshold value, $a>\psi_{\rm SR}$, the universality class of the short-range model is recovered. 
The critical exponents are determined solely by the decay exponent $a$ of the interaction strength, and, for a given $a$, they are universal for any distribution of the parameters. 
Concerning the asymptotic size-dependence of the average entanglement entropy of a block in the critical point, it is found to be logarithmic, similar to pure \cite{cc} and disordered \cite{rm} short-range quantum spin chains, but, as opposed to the latter systems, the prefactor is non-universal, i.e. disorder-dependent here.  

The results obtained in this work for the RTIM are relevant for other ferromagnetic quantum spin models, as well, such as the quantum Potts model \cite{sm}. 
In addition to this, we mention here a paradigmatic model of epidemic spreading, the {\it contact process} \cite{cp,liggett}, the absorbing phase transition of which, in the presence of disorder and for short-range interactions, falls into the same universality class as the RTIM, at least for strong enough initial disorder \cite{hiv}. 
The model studied in this work corresponds to a one-dimensional, disordered, long-range contact process, in which active sites become inactive with quenched random rates and the activity spreads to any other (inactive) site with rates decaying as a stretched exponential function of the distance. The SDRG decimation rules of this model being essentially identical to those of the RTIM, its absorbing phase transition is expected to be described by the IDFP found in this work, and the critical exponents characterizing the disorder-averaged quantities can all be expressed by $\psi$, $x$, and $\nu$ by the help of a scaling theory \cite{hiv}. For instance, the average survival probability, which is the probability that there is at least one active site at time $t$ if the process has been started from a single active site at $t=0$, decreases in the critical point asymptotically as $\overline{P(t)}\sim (\ln t)^{-\overline{\delta}}$ with 
$\overline{\delta}=x/\psi$.

We have seen that, as opposed to pure systems, where long-range interactions decaying faster than any power of the distance are usually irrelevant, 
for models with an infinite-disorder critical point, a stretched exponentially decaying interaction is able to alter the universality class of the 
transition. 
In higher dimensions $d>1$, the critical behavior of the short-range RTIM is still controlled by IDFPs with generalized critical exponents $\psi_d\approx 1/2$ 
varying weakly with $d$ \cite{ki_d}. 
On the basis of the results obtained in this work, we expect that a stretched exponentially decaying interaction with $a<\psi_d$ is relevant in any dimension and results in modified IDFPs with $a$-dependent critical exponents.  

\ack
This work was supported by the J\'anos Bolyai Research Scholarship of the
Hungarian Academy of Sciences, by the National Research Fund under grant no. K109577 and partially supported by 
the European Union and the
European Social Fund through project FuturICT.hu (grant no.:
TAMOP-4.2.2.C-11/1/KONV-2012-0013).


\end{document}